



\def \phead #1 {\bigskip \noindent {\bf #1 } \medskip } 

\def\bbhead#1{			
  \goodbreak\vskip 0.29truein	
  {\immediate\write16{#1}      
   \noindent {\bf #1}\par}
   \nobreak\vskip 0.26truein\nobreak}
     
\def\bhead#1{			
  \vskip 0.27truein		
  {\indent {\bf #1} \par}        
   \nobreak\vskip 0.16truein\nobreak}
     
\def\ihead#1{			
  \vskip 0.20truein		
  {\noindent {\it #1} \par}
   \nobreak\vskip 0.10truein\nobreak}






\def \half {{1\over 2}}

\def \pe {^\perp}
\def \la {\langle}
\def \ra {\rangle}

\def\pmm#1{\setbox0=\hbox{#1}
  \kern-0.2em\copy0\kern-\wd0
  \kern+0.2em\copy0
}
\def \veb {\bf}  


\def \b {{\bf b}}

\def \G {{\veb G}}

\def \h {{\veb h}}

 

\def \pp {{\veb p}}
\def \p {{\veb p}} 
\def \qq {{\veb q}}
\def \q {{\veb q}}
\def \Q {{\veb Q}}


\def \x {{\veb x}}
  
\def \RR {\rm {{\kern-0.2em I \kern-0.49em R}}}
\def \ZZ {\rm {{\kern-0.2em Z \kern-0.73em Z}}}




\def \z {{\bf z}}
\def \hbarr {{\overline {\h}}}
\def \Kbar {{\overline {K}}}
\def \zB {{z_{B}}}
\def \zW {{z_{W}}}
\def \hW {{h_{W}}}
\def \hB {{h_{B}}}
\def \Ksix {{K_1}}
\def \Kc {{K_2}}
\def \zsix {{z_1}}
\def \zc {{z_2}}
\def \hsix {{h_1}}
\def \hc {{h_2}}
\def \Kp {{K_+}}
\def \Km {{K_-}}
\def \pp {{\bf p}}
\def \tz {{\tilde \z}}
\def \O {{\bf O}}
\def \y {{\bf y}}

\def \rem #1 {{\sl #1}}

\noindent {To appear {\it J. Stat. Phys.}, February, 1997.}

\bigskip
\centerline{\bf NEW TWO-COLOR DIMER MODELS WITH CRITICAL 
GROUND STATES  }

\centerline 
{R. Raghavan, 
Christopher L. Henley,\footnote *
{To whom correspondence should be addressed}
Scott L. Arouh
\footnote \dag
{Current  address: Dept. of Physics, 
Univ. of California San Diego, La Jolla CA 92037}}

\centerline 
{Laboratory of Atomic and Solid State Physics}
\centerline 
{Ithaca, NY 14853-2501}
\bigskip \bigskip
Received
\bigskip \bigskip
\bigskip

{\it ABSTRACT:}
We define two new models on the square lattice, in which
each allowed configuration is a superposition of a 
covering by ``white'' dimers and one by ``black'' dimers. 
Each model maps to a solid-on-solid (SOS) model in which
the ``height'' field is two dimensional. 
Measuring the stiffness of the SOS fluctuations in the rough
phase provides critical exponents of the dimer models. 
Using this ``height'' representation, we have performed
Monte Carlo simulations.
They confirm that each dimer model has critical correlations
and belongs to a new universality class.
In the ``dimer-loop'' model
(which maps to a loop model)
one height component is smooth but has unusual
correlated fluctuations; the other height component is rough.
In the ``noncrossing-dimer'' model, the heights are 
rough having two different elastic constants; 
an unusual form of its elastic theory implies anisotropic critical correlations.

\medskip

\def \pe {^\perp}

\bbhead {1. INTRODUCTION}

Certain discrete spin models 
are defined by a constraint on the configurations, 
with all allowed configurations receiving equal statistical weights.
Among these are dimer models and  ice models, as well as the
(highly degenerate) ground state ensembles  of antiferromagnetic Potts models
and frustrated Ising models.
Many of the models on two-dimensional lattices
are critical, in that correlation functions  decay with distance
as power laws [1]; and every one of these
critical models has been found to have a ``height representation'' [2,3], 
that is, a 1-to-1 mapping of the microstates to those of a
sort of interface model.

A ``height representation'' means that
each configuration can be mapped to a configuration of ``heights''
$z(\x)$ on the lattice, i.e. to a configuration of a sort of 
solid-on-solid (SOS) model. Furthermore, this SOS model is (usually)
in a rough phase, described at long
wavelengths by a gradient-squared elastic free energy. From this we
follow a route familiar in the context of the two-dimensional XY model [4]:
the height fluctuations diverge logarithmically.
Furthermore, local spin operators can be represented as 
complex exponentials of the height variables. 
Hence the correlations of those operators
decay as the exponential of a logarithm, explaining the criticality
and providing formulas for the critical exponents in terms
of the elastic constants [5].
(See Sec. 2.1 and 2.5, below, for more about this derivation). 
The height approach also yields the critical exponents
of the specific heat and correlation length as $T\to 0$ in 
models where the excitations map to topological defects
of the interface model.

It is possible for the ``height'' variable 
$\z(\x)$ to have dimensionality $d\pe  > 1$, 
so that each configuration
corresponds to a $2$-dimensional surface embedded
in a $(2+d\pe)$-dimensional hyperspace. 
After this point was noticed [3,6], it was a natural step
to construct new models which admit
height mappings, in the hopes that the heights would be ``rough'' and 
the corresponding spin models would be critical.
Since only one model with $d\pe>1$ was solved in the past
(the 3-coloring of the honeycomb lattice [6,7,8]),
it is likely that new models of this type 
belong to new universality classes.
For example, Read [9] conjectured that the 4-coloring of a square lattice
(which has $d\pe=3$)
would belong to the universality class of the SU(4) Wess-Zumino-Witten
field theory; this has been supported by Monte Carlo simulations [10].
The triangular Ising antiferromagnet, for any spin number 
$S>1/2$, is a height model with $S$-dependent exponents;
in this case $d\pe=1$ [11].
(The constrained 4-state Potts antiferromagnet on the square lattice,
a case where $d\pe=5$, was also studied, 
but in that model the height field appears to be smooth [12].)
In some special cases the ``height''approach allows the exact determination
of an exponent [2,13]. 

The ``height model'' approach is 
closely similar to the ``Coulomb gas'' theory
of critical models [14]. There are three differences in emphasis:
  \par \qquad (i) Our ``height space'' may have more than one dimension.
(In Coulomb gas language, there may be more than one flavor of ``charge''.)
  \par \qquad (ii) We limit ourselves to mappings that are one-to-one between 
the microscopic spin configurations and 
the microscopic height configurations, modulo some global arbitrary
choices (such as a constant shift of all the heights);
the Coulomb gas picture is often applied to less trivial 
transformations of the partition function,  in which additional 
variables are introduced (and the original variables may be summed
over, as in a duality transformation).
  \par \qquad (iii) Because the mappings are one-to-one, the height 
representation can be directly used to analyze Monte Carlo simulations.

Here, we report two models representing two new universality classes. 
Each model is built from two (interacting) copies of the dimer covering 
of a square lattice, and each corresponds to $d\pe=2$. 

\bhead {1.1 Definitions of the models}

The building block for both of our models
is the complete covering of a square lattice  by
dimers (i.e. each vertex has one end of a dimer), as in
Fig. 1(a), 
which will be called the ``simple dimer model'' in this paper.
It is a height model with $d\pe=1$, and
the elastic constant  is known since the
model is exactly solved [15].
The new models are:
 
\par \qquad (i) 
``Dimer-loop model''. Each allowed configuration consists
of ``black'' dimers, forming a complete dimer covering of  the
square lattice, and ``white'' dimers forming another complete dimer
covering. We exclude any configurations in which the same bond is
occupied by both black and a white dimer; thus the dimers form
non-intersecting loops covering all vertices, 
each loop consisting of alternating white and black dimers
(see Fig. 1(b)). 

\par \qquad (ii) 
``Noncrossing dimer model''. 
Each allowed configuration consists of 
``black'' dimers, forming a complete dimer covering of  the
square lattice, and ``white'' dimers forming a complete dimer
covering its {\it dual} lattice. 
Now, we exclude any configurations in which  a white dimer
crosses a black dimer (see Fig. 1(d)). 

In either model, 
it can be checked immediately that the interaction between colors is not
so strong as to make the models trivial: that is, 
given a typical configuration of (say) black dimers, 
there is still an  extensive entropy of allowed ways to arrange
the white dimers. 

\ihead {Mappings of the dimer-loop model}


As its name implies, the dimer-loop model
may also be viewed as a loop model, in a family  which also
contains the equal-weighted six-vertex model.
This is useful, firstly because it makes contact with the large
literature on loop models and secondly because it suggests
concrete interpretations of the two components of the height
space (see Sec. 2.3, below).

Take each microstate in this model and make all dimers the
same color. The new ensemble contains all ways to cover
the sites with nonintersecting loops, 
each site being covered by exactly  one loop, 
with a fugacity $n=2$ for each loop
(from the two ways in which it could have been colored). 
This is the square lattice version of the
``fully-packed loop''  (FPL) model, which was originally defined on the
honeycomb lattice [7]. 
The square lattice FPL model is related
to the dimer-loop model in much the same way that the honeycomb FPL model
is related to the honeycomb 3-coloring  model [8].
The case $n=1$ of the square lattice FPL model
is equivalent to the 6-vertex (square ice) model with 
every configuration weighted equally [16]. 
The logic of [7] suggests that the dimer-loop model maps
to some $O(n)$ model;
it would not be surprising if our case ($n=2$) were exactly soluble and
had rational exponents.

\bhead  {1.2 Outline of the paper}

In Sec. 2 (see also the Appendix), we review the basic notions
of the theory of height models, as applied to the two-color 
dimer models: the height representation itself,
the special microstates called ``ideal states'' which are
flat in the height representation, 
the elastic theory, and finally the calculation of critical exponents.
Sec. 3 describes the Monte Carlo simulations of both
of the two-color  dimer models; in addition, we simulated the
exactly solved simple dimer model and the BCSOS model as tests.
The results are collected in Sec. 4: 
we found that in the ``dimer-loop model'', 
one component of height space remains rough, 
while the other component becomes smooth (but with anomalous
correlations suggesting a mediated interaction);
in the ``noncrossing-dimer'' model, 
both components of height space are rough.
Sec. 5 contains a summary, some remarks on 
the new Monte Carlo methods introduced in this work 
(and Refs. [10-13]), and speculations on possible analytic
extensions of this work. 

\bbhead {2. HEIGHT REPRESENTATIONS}

This section contains the theoretical apparatus for 
describing the two-color dimer models. 
As a motivation, 
it begins by reviewing how all critical exponents can 
be derived from the elastic constants, in a  generic height model. 
The rest of the section works out the necessary modification
for the special cases of the two-color dimer models, with 
the complications of a two-dimensional height space.

\bhead {2.1 Fluctuations and correlation functions in height models}

Consider a generic height model with (for now) one
component in the height space. 
The microscopic heights 
$\{ z(\x) \}$ are defined on each lattice site, 
and they are coarse grained so as to define a
continuous height field $h(\x)$.
As noted above, the height fluctuations are usually ``rough'';
here this will mean that the fluctuations of $h(\x)$
are weighted according to a gradient-squared elasticity, 
$F = \int d^2 \x f(\nabla h(\x))$, with a free energy density
   $$f(\nabla h)  = \half K |\nabla h|^2  .
   \eqno (2.1)$$
Inserting (2.1) into a Fourier transform gives
  $$F = \sum _\pp \half K|\pp|^2 |h(\pp)|^2. 
   \eqno (2.2)$$
Hence by equipartition,
   $$ \la |h(\pp)|^2 \ra = {1\over {K|\pp|^2}}.   
   \eqno (2.3)$$
 From (2.3) we can calculate the height correlation function 
$C(\x) \equiv \la [h(\x)-h(0)]^2\ra$ , 
   $$C(\x) 
    = \int {{d^2\pp }\over{(2\pi)^2}}
     {{ 2 (1-\cos (\pp\cdot\x)) } \over {K |\pp|^2  }} .
   \eqno (2.4) $$
   $$ \qquad \simeq
     {\rm const}+ {1\over{\pi K}} \ln |\x| .
     \eqno (2.5)$$

Using (2.5), we can infer the correlation functions
of any operator $O(\x)$ which is a local function of the spins
(or dimers) in the vicinity of $\x$. 
First, as will be justified in Sec. 2.2, 
we can write $O(\x)\approx {\tilde O}(h(\x))$
where ${\tilde O}(h)$ is a periodic function dominated by 
   $${\tilde O}(h) \sim \exp(\pm i G\pe h(\x))
   \eqno(2.6)$$
Then 
     $$\la O(\x) O(\O)\ra  \sim {\rm Re} \la \exp(iX) \ra$$
where $X=G\pe(h(\x)-h(\O))$.  
But $\la \exp(i X) \ra =  \exp(-\half \la X^2\ra ) $, since
(2.2) implies $X$ is a Gaussian random variable 
(more precisely, the long-wavelength contribution is). 
Hence
     $$\la O(\x) O(\O)\ra 
     \sim \exp(-\half G^2 C(\x))
    \sim r^{-\eta(G\pe)},  
    \eqno (2.7)$$
where
    $$\eta(G\pe) = { 1 \over{2\pi K}} { {G\pe}^2 }
    \eqno (2.8)$$

In some models, we can admit a small density of topological defect
excitations characterized by a Burgers vector $b$. 
(That is, if we follow the value of $h(\x)$ along a 
path circling the defect clockwise, then $\h(\x)$ is not well-defined
but has a net change of $b$.)
The correlation between the defects and antidefects define
another kind of correlation function, with ``vortex'' exponents given by
    $$\eta_v(b) = { K \over{2\pi }} { {b}^2 }
    \eqno (2.9)$$
In the case of dimer models, 
the defects are uncovered (or multiply covered) sites.

Thus, if we know the value of $K$  and the allowed values
of $G\pe$ and $b$, then eqs. (2.8) and (2.9)
provide all the critical exponents that may occur in the model.
This was worked out in detail for the case of the
triangular Ising antiferromagnet by Bl\"ote, Hilhorst, and Nienhuis [5]. 

\bhead {2.2 Ideal states}

It is convenient to analyze height models by identifying special
``ideal states'', which are microstates in which the
configurations are periodic in real space [2,3,10,11,12].
So far, this notion does not have any rigorous basis, 
but it is a useful shortcut for identifying the 
``repeat lattice'' in height space. The possible Burgers
vectors $\b$ of defects (entering eq. (2.9))
belong to the repeat lattice, 
while the allowed wavevectors $\G\pe$ (entering eq. (2.8))
of the operators belong to its reciprocal lattice. 
Hence knowing the repeat lattice and the elastic constants
means knowing the possible critical exponents of the model.

Ideal states can be identified by either of the following criteria:
   \par\qquad 
(i). Flatness: the variance of $\{ \z(\x) \}$ within the ideal state 
is minimal.\footnote *
{Note that no microstate has $\z(\x)={\it constant}$, as will be
clear from the form of the height maps, e.g. (2.11) or (2.14-2.15).} 
   \par\qquad 
(ii). Entropy: define two microstates to be ``neighbors'' in
configuration space  if they differ from each other on 
the minimum number of sites.
(Such minimum differences are shown in Fig.~2). 
Then an ideal state is one which has the maximum number of neighboring
microstates. (That is, of all configurations, the ideal states have
the maximum density of sites at which a minimal
rearrangement can be performed.)
Typically, the same microstates satisfy both criteria. 

Each height model has several degenerate ideal states
related by symmetry operations.
Every ideal state is associated with a ``height'' $\hbarr$, the spatial
average of $\z(\x)$ over its $2\times 2$ unit cell.
An arbitrary microstate can thus be divided into a set of 
domains  of symmetry-related ideal states, with each 
domain considered  to have a uniform value of $\hbarr$. 
This is the intermediate step in the coarse-graining process, where we
are already treating $\x$ as a continuous variable, 
but $\hbarr$ takes on discrete values.
In the final step of coarse-graining, we replace $\hbarr$ by $\h(\x)$ which 
takes on a continuum of values. 

There is a finite number of ideal states, but the values of $\z$ 
are unbounded; for each allowed $\hbarr$ value there is unique ideal state, 
but for a given ideal state the
the possible $\hbarr$ value 
is only fixed modulo a Bravais lattice called the ``repeat lattice''.
Thus the pattern of dimers near any given site
is a function of $\hbarr$, and hence
any local operator $O(\x)$ can be expressed as a function $\tilde O (\hbarr)$;
furthermore $\tilde O(\hbarr)$ must have the periodicity of the repeat lattice;
Thus it can be written as a Fourier sum of terms 
of form $\exp(i\G\pe\cdot \h(\x))$ where $\G\pe$ are
reciprocal lattice vectors of the repeat lattice. 

The possible values of $\hbarr$ in height space
constitute the ``ideal-state graph''.
The fact that the local configuration is more likely
to be near an ``ideal state'' than far from one, 
is expressed by an additional term $V(\h)$ which should be added 
to the free energy density (2.1). 
This $V(\h)$ can be called a ``locking term'' since 
if $V(\h)$ were
strong enough, it would force the system into long-range order
in one of the ideal states.
Just like the local operators mentioned above, 
$V(\h)$ has the periodicity of the ideal-state graph,
and it too can be labeled by a reciprocal lattice vector
of the repeat lattice, $\G_{lock}$. 
Standard Kosterlitz-Thouless theory [4] predicts that $V(\h)$ 
becomes relevant in the renormalization-group sense when 
    $$\eta(\G_{lock})< 4,
    \eqno (2.10)$$
At that point the height component ought
to become smooth (a roughening or ``locking'' transition).
Notice that if the heights are in a ``smooth'' phase, then 
the reasoning in Sec. 2.1 implies that the correlations decay to a fixed
nonzero value, i.e.  that the spin model
has long-range order.

\bhead {2.3 Height mappings for two-color dimer models}

In general height mappings,  one defines a ``microscopic'', discrete-valued
height function $\z(\x)$ such that the step in $\z(\x)$ between adjacent sites
is a function of the adjacent spins (or, in this case, dimers). 
The height representations for two-color models are built in
an obvious fashion from those of the simple dimer model. 

\ihead {2.3.1 Review of simple dimer model}

We define heights $z(\x)$ on the {\it dual} lattice, i.e. in 
the centers of each plaquette. 
We take the standard orientation of every edge pointing 
from the even to the odd vertex. 
Say that $\x$ and $\x'$ are neighboring plaquette centers, such that
the edge between them is oriented left to right (when looking from 
$\x$ to $\x'$); then define the height difference to be
    $$z(\x')-z(\x) = +1(-3)  \hbox {\ if there is no dimer (is a dimer)}
   \eqno (2.11)$$
on the edge  between $\x$ and $\x'$. 
Traversing the four plaquette centers surrounding any vertex,
the total change in $z(\x)$
is $1+1+1-3=0$, since every vertex of the original lattice has exactly
one dimer emanating from it. Thus $z(\x)$ is well-defined everywhere, 
{\it provided the dimer covering is complete.} 
The $\{ z(\x) \}$ values are indicated in Fig. 1(a). 

It should be noted that 
the triangular Ising antiferromagnet ground states are (essentially)
equivalent to the dimer coverings of the honeycomb lattice
(each violated bond corresponds to a dimer) 
so the construction of $z(\x)$ here is closely analogous to that of [5],
and indeed the square and honeycomb dimer coverings have
the same critical exponents [15]. 
The height map (2.11) for the square lattice case was,
in fact, discovered several times in the context
of quantum dimer models and nearest-neighbor valence-bond
ground states of $s=1/2$ antiferromagnets [17]; however, 
it has never been applied to derive exponents 
in the spirit of Bl\"ote, Hilhorst, and Nienhuis [5]. 

\ihead {2.3.2 Height map of the dimer-loop model}

Clearly, we can construct heights for the two-color models
by letting $\z= (\zB, \zW)$ where $\zB$  and $\zW$ are the height
configurations of the black and white dimer configurations, 
constructed by the rule given in (2.11). 

Let's define the even combination of heights 
    $$\zsix \equiv  \half (\zB+\zW)
     \eqno(2.12) $$
and the odd combination of heights
    $$\zc = {1\over 4} (-\zB+\zW)
     \eqno(2.13) $$
Under the exchange ${\rm black} \leftrightarrow {\rm white}$, 
$(\zsix, \zc) \to (\zsix, -\zc)$.

By inserting (2.12) and (2.13) into (2.11), we can write
the dimer-loop height rule in components.
When the edge between $\x$ and $\x'$ is
oriented as in (2.11),
    $$\zsix(\x')-\zsix(\x) = +1 (-1),  
     \hbox {\quad if there is no dimer (is a  dimer) }
    \eqno (2.14)$$
on that edge, and
    $$\zc_(\x')-\zc(\x) = 0 (+1,-1)  
     \hbox {\quad if there is no dimer (black dimer, white dimer)} 
    \eqno (2.15)$$
on the edge. 

As already mentioned in Sec. 1.1, 
if we do not distinguish black from white,
this model reduces to the 
fully-packed loop (FPL) model with loop fugacity $n=2$.
The FPL model for general fugacities admits the same configurations, 
but with different weights.
Any FPL configuration can be mapped 1-to-1 to a configuration of the 
six-vertex (or ``ice'') model:\footnote *
{A six-vertex configuration means a pattern of arrows
on the dual lattice such that two ice-arrows are incoming and two are
outgoing at every vertex.} 
a bond receives an ice-arrow pointing from the even to the odd vertex, 
if that bond is occupied by loop segments, 
and pointing oppositely if the bond is vacant.  

The six-vertex model has a well-known height mapping,  
indeed its height configurations $\{ \zsix(\x) \}$
are simply the microstates of the body-centered solid-on-solid 
(BCSOS) model [18].
The $\{ \zsix (\x) \}$ sit on the dual lattice and are defined by 
$\zsix(\x')-\zsix(\x) = +1(-1)$, 
according to whether the ice-arrow (viewed looking from $\x$ to $\x'$)
points rightwards (leftwards) along the edge between $\x$ and $\x'$.
It can easily be checked that the $\zsix$ configuration defined in 
(2.14) for the
dimer-loop model is identical to the $\zsix$ configuration defined 
definted by mapping to the FPL and then to the 
six-vertex model, followed by the six-vertex height rule. 
(See Fig. 1). 

The component $\zc$ has its own simple interpretation as an SOS model.
The loops are simply {\it contours} of equal height $\zc$.
If one switched the colors ${\rm white} \leftrightarrow {\rm black}$ 
within one loop, then a step up would be turned into a step down. 
We could view the rule for $\{ \zc(\x) \}$ as a constrained
SOS model: besides the standard 
restriction that $|\zc(\x')-\zc(\x)|\leq 1$, we also require that
every block of four plaquettes must include exactly two pairs 
of neighboring plaquettes on which $\zc(\x')=\zc(\x)$. 
It is amusing
that this SOS model, which would appear to be described by only one height 
variable, is one-to-one equivalent to the original 
two-color dimer-loop model 
and thus contains $\zsix(\x)$ as  a second, hidden  height variable.

\ihead {2.3.3. Height map of the noncrossing-dimer model}

To define the heights for both colors of dimer
in the noncrossing-dimer model, we arbitrarily define
the dual lattice site at $\x+(\half,\half)$ to have the same
parity as the original lattice site at $\x$.
Then we construct height configurations $\zB(\x)$ and $\zW(\x)$ 
as before. However, in this model the $\{ \zB(\x) \}$ live on the original
lattice while 
the $\{ \zW(\x) \}$ live on the dual lattice, so we cannot at this point
define even and odd combinations like (2.12)-(2.13);
we shall do so later on, at the coarse-grained stage (see Sec. 2.4). 

\ihead {2.3.4 Ideal states of two-color dimer models}

The proposed ideal states for the models in this paper
are shown in Fig.~3; they all have
$2\times 2$ unit cells. 
The black or white dimer parts of these
two-color ideal states are just ideal states of the simple dimer model.
It is straightforward to identify ideal states
for the simple dimer and noncrossing-dimer models, but it
is not so clear whether Fig.~3(b) or Fig.~3(c) is
the best choice in the case of the  dimer-loop model.
We prefer Fig.~3(b) as it has a smaller variance of the $\zsix$ component;
indeed the $\zsix(\x)$ pattern here is the same as the ideal state
of the BCSOS model. On the other hand, 
the alternative Fig.~3(c) does have a smaller variance of the $\zc$ component, 
which is found to be in a ``smooth'' phase as
will be elaborated in Sec.~4

Fig. 1(a) illustrates that real configurations
can indeed be broken into ``ideal state'' domains. 
Note how all the  sites with height ``1'', and most of those
with height ``0'' or ``2'', are unchanged from a certain ideal state;
this is not surprising
since such a small system has no room for a large height fluctuation.
On the other hand, the local patterns {\it appear} different from
those in an ideal state, 
in that one rarely finds more than two parallel dimers in a row.

Fig.~4 shows the ideal-state graphs for these models. 
For the case of the simple dimer covering, the 
$\hbarr$ values differ by $\pm 1$ between 
ideal states, and the repeat lattice constant is 4 
(corresponding to the four symmetry-related ideal states). 
It turns out that the two-color ideal states are made from
combinations of the simple-dimer ideal states, but not every
combination is ideal or is optimal. 
For the dimer-loop case, Fig.~4(b), each black dimer ideal state
can be combined with only one white dimer ideal state so there
are only 4 ideal states.
For the noncrossing-dimer case, Fig.~4(d), 
simple-dimer states can be combined only if both colors of dimers
are oriented the same way, so there are 8 ideal states. 

\bhead {2.4 Elastic theory}

Next we will work out the allowed symmetry form
of the gradient-squared elastic free energy density $f(\nabla \h)$. 
It is quite possible that a 2-dimensional height space
would have only one elastic constant 
(as for the honeycomb two-coloring [6,8]). 
However, each two-color dimer model here turns out
to have two distinct elastic constants. 

We proceed in the spirit of Landau theory, 
by first identifying the tranformations of $\{ \z(\x)\}$
which are induced by a given lattice symmetry operation.
Next we identify the coarse-grained version of this
symmetry operation, and finally we require that 
$f(\nabla \h)$ be invariant under this symmetry. 
The elastic theory will be related
to critical exponents in Sec.~2.5.

\ihead {2.4.1 Simple dimer model.}

The microscopic transformations are

\qquad (i). $x_1 \to x_1+1$ induces $z(\x) \to -z(\x)$ 
(because our standard orientation of edges alternates
between even and odd)

\qquad (ii). $x_1 \to -x_1$ (mirror line through a row of lattice points) 
induces $z(\x) \to -z(\x)$, and of course $\nabla_1 \to -\nabla_1$
(we take $\nabla_i \equiv \partial/\partial x_i$).
On the other hand, a mirror line through a row of plaquette
centers is represented by $x_1\to 1-x_1$ and induces $z(\x) \to -z(\x)$. 

\qquad (iii). A 4-fold rotation about a vertex induces $z(\x) \to -z(\x)$, 
$\nabla_1 \to \nabla_2$, $\nabla_2 \to -\nabla_1$. 
On the other hand, a 4-fold rotation about the center of 
a plaquette induces $z(\x)\to z(\x)$.

Now we consider all possible terms bilinear in gradients of $h(\x)$. 
By (iii) $\nabla_1 h \nabla_2 h \to - \nabla_1 h \nabla_2 h$
under a symmetry,
so this term must have zero coefficient in the free energy.
Also $(\nabla_1 h)^2  \to (\nabla _2 h)^2$, so that those two terms must
have equal coefficients. 
Thus  the generic form is
    $$f(\nabla h) = \half K [(\nabla_1 h)^2 +(\nabla_2 h)^2]
   \eqno (2.16)$$
    
\ihead {2.4.2 Dimer-loop model}

In the two-color dimer models, the above reasoning still holds for
all terms which depend purely on $\hW$ or on $\hB$
(the coarse-grained $\zW$ and $\zB$ components), thus we
only need to check the $\hB$-$\hW$ cross-terms. 

In the dimer-loop model, the reflection symmetry (ii) transforms
$\nabla_1 \hB \nabla_2 \hW \to - \nabla_1 \hB \nabla_2 \hW$, 
thus such terms must be absent in $f(\nabla \h)$. On the other hand, 
no symmetry excludes the terms
$\nabla_1 \hB \nabla_1 \hW$ and $\nabla_2 \hB \nabla_2 \hW$, 
and the 4-fold rotation symmetry merely demands that 
their coefficients  be equal. 
After diagonalizing the quadratic form in $\nabla \h$ we obtain
    $$f (\nabla \h) = \half  \Ksix |\nabla\hsix|^2
    + \half  \Kc |\nabla\hc|^2 
   \eqno (2.17)$$
where 
    $$(\hsix,\hc) = (\half (\hB+\hW), {1\over 4} (\hB-\hW))
     \eqno (2.18)$$
is the coarse-grained version
of $(\zsix, \zc)$ defined by  (2.12) and (2.13). 

We might have guessed the form (2.17) from a glance at the ideal-state graph
(Fig.~4(b)), which shows
that the $\hsix$ and $\hc$ directions are not equivalent by any symmetry.
Fig.~4(b) also suggests that the stiffness should be greater for the 
$(1,-1)$ projection of height space than for the $(1,1)$ projection, 
since fluctuations in the former direction must pass through several 
steps to get from one favorable (ideal) state to the next one.
After the above changes of variables, this expectation
translates to the inequality $\Kc \ge 4 \Ksix$. 

\ihead {2.4.3 Noncrossing-dimer model}

This case is subtler. 
A mirror line that passes through a lattice point in the 
{\it black} lattice runs through {\it plaquette} centers
in the dual {\it white} lattice, hence a reflection induces
e.g. $\nabla_1 \to -\nabla_1$, $\hB \to \hB$, $\hW\to -\hW$;
the analogous thing happens with the 4-fold rotation. 
For this case, the cross-terms
$\nabla_1 \hB \nabla_1 \hW$ and $\nabla_2 \hB \nabla_2 \hW$, 
are excluded, while
$\nabla_1 \hB \nabla_2 \hW$ and $\nabla_1 \hB \nabla_2 \hW$
are allowed, with the 4-fold symmetry demanding that their
coefficients be equal. 

We adopt height-space coordinates rotated by $45^\circ$, 
    $$h_1\equiv 
   {1\over \sqrt2}
    (\hB + \hW), \qquad
    h_2\equiv 
   {1\over \sqrt2}
(-\hB + \hW)
    \eqno (2.19)$$
-- as in the dimer-loop model, except that here $h_1$ and $h_2$ 
are related by a symmetry.
To represent the elasticity in the simplest form, we
must also rotate {\it real} space by $45^\circ$ defining 
coordinates $\y$:
    $$y_1\equiv 
   {1\over \sqrt2}
(x_1 + x_2), \qquad
    y_2\equiv 
   {1\over \sqrt2}
(-x_1 + x_2).
    \eqno (2.20)$$
and the wavevector $\qq$ is defined by the corresponding rotation 
on $\pp$. 
Under these changes of variables, the elastic terms noted in the first 
paragraph yield as the generic form,
\def \naby#1 {{{\partial}\over{\partial y_#1}}}
     $$f(\nabla\h) = \half \Kp \big( |\naby1 h_1|^2 + |\naby2 h_2|^2 \big)
     + \half \Km \big( |\naby1 h_2|^2 + |\naby2 h_1|^2 \big)
      \eqno (2.21)$$
with two independent elastic contants $\Kp$ and $\Km$.

A consequence of (2.21) plus equipartition is that the generalization
of (2.3) reads
   $$\la |h_1(\pp)|^2 \ra = \big(\Kp q_1^2 + \Km q_2 ^2 \big) ^{-1}
   \eqno (2.22) $$
with $\la |h_2(\pp)|^2 \ra$
defined by exchanging $1\leftrightarrow 2$ in (2.22).

There is a simple argument that $\Kp>\Km$. 
It can be checked that, with our even/odd conventions, 
the contribution to $\partial \hB/\partial x_1$
made by a black dimer on a vertical bond
has the same sign as the contribution to 
$\partial \hW/\partial x_2$  that a white dimer on the
horizontal crossing bond  would have made.
The exclusion between these possibilities has the effect 
of a positive term in the free energy proportional to 
$(\partial \hB/\partial x_1)(\partial \hW/\partial x_2)$;
a similar argument works for $x_1 \leftrightarrow x_2$. 
After we carry out the above changes of variable, we find that
$\Kp-\Km$ in (2.21) is proportional to the positive coefficient of those terms. 


\bhead {2.5 Critical Exponents for two-color dimer models}

We can now calculate the correlation function exponents 
by the path sketched in Sec.~2.1, but adapting to 
$d\pe >1$ in two ways:
\par\qquad 
(i) the elasticity 
has a more complicated form than eq. (2.1),
when there is more than one elastic constant.
\par\qquad 
(ii) the spin operators are now represented by 
   $$\exp(i \G\pe \cdot \h(\x))
   \eqno (2.23)$$
where $\G\pe$ is a vector of the reciprocal lattice
of the ``repeat lattice'' in the height space;\footnote * {
The dominant wavevector $\G\pe$  for a given operator 
can be deduced easily with the aid of the ``ideal states'' graph.
We have omitted this detail in the present paper;
the method is explained in Ref.~[10].}
similarly, a defect Burgers vector $\b$ must be a vector of the
repeat lattice. 

The previously known example of a two-dimensional height space was
the 3-coloring of the honeycomb lattice. 
That model is (essentially) equivalent to the 
3-state-Potts antiferromagnetic ground states on the Kagome lattice [6], 
the 4-state-Potts antiferromagnetic ground states on the 
triangular lattice  [3], and the ``fully-packed loop (FPL) model''  
on the honeycomb lattice [7,8]. 
Given one previously known exponent of this model, Huse and Rutenberg  
used the arguments reproduced in Sec.~2.1 
to find the correlation function exponent
of the Kagom\'e 3-state-Potts model [6] (later additional exponents
in the FPL model were calculated in the same fashion [8]). 

\ihead {2.5.1 Dimer-loop model}

In this model, dimer-dimer correlations which do not depend on
the dimer color are governed by $\G\pe =(G\pe_1,G\pe_2)=(2\pi/2, 0)$; 
those that
treat black and white dimers as having opposite signs are 
governed by $\G\pe=(2\pi/4, 2\pi/2)$. 
The exponents are given  by
    $$\eta(\G\pe)={1\over{2\pi K_1}} |G\pe_1|^2
    +{1\over{2\pi K_2}} |G\pe_2|^2
    \eqno(2.24)$$

A defect of type $\b=(b_1,b_2)=(2,1)$ 
corresponds to an endpoint of a loop, while 
$\b=(0,1)$ corresponds to having
two successive segments of the same color in a loop.
The exponents are given by
   $$\eta_v(\b)= {{K_q}\over{2\pi}} |b_1|^2
    +{{K_2}\over{2\pi}} |b_2|^2
    \eqno(2.25)$$

\ihead {2.5.2 Noncrossing-dimer model}

In this case, there are nontrivial modifications in the derivation
on account of the anisotropic elasticity. 
The height correlation function is
$C_k(\x) = \la [h_k(\x)-h_k(0)]^2\ra$ (for $k=1,2$). 
Substituting from (2.22), and rescaling the coordinates
     $$(q'_1, q'_2) \equiv (\lambda q_1, \lambda^{-1} q_2), 
     \qquad (y'_1, y'_2) \equiv (\lambda^{-1} y_1, \lambda y_2) 
      \eqno (2.26)$$
(where $\lambda \equiv (\Kp/\Km)^{1/4}$), 
we obtain
   $$C_1(\x) =  
    \int {{dq'_1 dq'_2}\over{(2\pi)^2}}
     {{ 2 (1-\cos (\q'\cdot \y')) } \over { \Kbar |\q'|^2}}
     = {\rm const}+ {1\over{\Kbar \pi}} \ln |\y'| ,
     \eqno (2.27)$$
where $\Kbar \equiv (\Kp\Km)^{1/2}$.
($C_2(\x)$ is also given by (2.27) but with 
$\lambda \leftrightarrow \lambda^{-1}$ in  (2.26)). 
Recall that in all these formulas, $\y'$ is rewriting of $\x$ through
(2.20); in particular
if we write $\x=|\x|(\cos \theta, \sin \theta)$, then
$|\y'| = |\x| f(\theta,\lambda)^{1/2}$
where 
   $$f(\theta)=  \lambda^2 \cos^2(\theta-\pi/4) + \lambda^{-2}  
        \sin^2(\theta-\pi/4). 
     \eqno (2.28)$$ 
Thus if an operator is written in the form
   $O(\x)\sim \exp (i\G\pe\cdot \h(\x))$, 
it follows as usual that
   $$\la O(\x)O(0)\ra = \exp (-\half \la[\G\pe\cdot (\h(\x)-\h(0))]^2 \ra ). 
   \eqno (2.29)$$
This can be written 
   $$\la O(\x)O(0)\ra = \exp (-\half [(G_1\pe)^2 C_1(\x) +
   (G_2\pe)^2 C_2(\x)] $$
where $\G\pe$ has been resolved into components in the $(h_1,h_2)$ basis. 
Inserting the logarithm (2.27) into the exponential (2.29), 
we finally get
   $$\la O(\x)O(0)\ra \sim
    |\x|^{-\eta(\G\pe)} 
   f(\theta)^{\eta_1} f(\theta-\pi/2)^{\eta_2}
   \eqno (2.30)$$
where 
   $$\eta_k \equiv  {1\over{2\pi \Kbar}} (G_k\pe)^2$$ 
for $k=1,2$,  and 
   $$\eta \equiv  \eta_1 + \eta_2 
   =  {1\over{2\pi \Kbar}} |\G\pe|^2
    \eqno(2.31)$$ 

Thus, the exponents $\eta(\G\pe)$ 
for the noncrossing-dimer model, 
have just the form that
would follow from an isotropic effective elasticity with one
elastic constant $\Kbar$. 
However, the decay of correlations is not isotropic, as in 
other known cases, but has the
anisotropy factors $f()$ displayed in (2.30). 

The defect-defect correlation function corresponding to a
Burgers vector $\b$ would have an exponent
    $$\eta_v(\b)= (2\pi)^{-1}\Kbar|\b|^2
   \eqno (2.32)$$
but with exotic anisotropy factors analogous to  those in (2.30).

\bbhead {3. MONTE CARLO SIMULATIONS AND RESULTS}

Monte Carlo simulations were performed for the dimer-loop
and noncrossing dimer models
in square lattices with periodic boundary conditions. 
Besides the new models introduced in this paper, we have also 
simulated two exactly solved models as checks (both for debugging
of our simulation codes, and to test how much accuracy may be obtained from
this way of analyzing the results.)
(i) Since both $\zsix$ and $\zc$ have configurations like those of a 
simple SOS model, we chose to simulate the BCSOS model.
(ii) We also simulated the simple dimer model. 

In this section, we present in turn the update moves, 
the simulation protocol, and the numerical results.

\bhead {3.1 Update moves}

In some ``height'' models -- e.g. 
the square lattice four-coloring model [10] 
or the constrained Potts antiferromagnet [12] --
a nonlocal cluster or loop update move is required. 
In each of the two-color dimer models, however, a local update move, 
based on the rearrangements shown in Fig.~2,  was adequate. 
A ``pass'' thus consisted of visiting each site once (in 
random order), testing whether the site could be rearranged, 
and if so changing it. 
These update rules
satisfy detailed balance for the ensemble in which
each microstate receives equal statistical weight.

There is an important technicality associated with the question of
ergodicity, in the presence of our periodic boundary conditions.
Let $\Delta_1 \z = \z(L,0)-\z(0,0)$
and $\Delta_2 \z = \z(0,L)-\z(0,0)$. These are Burgers vectors
associated with the topologically nontrivial loops on the torus, and 
so they must belong to the repeat lattice.
Then $\Delta_1\z$ and $\Delta_2\z$ 
are conserved under all {\it local} update moves,
including ours. 
The average of the height gradient $\nabla_i \h$ is $\Delta_i \h/L$, 
thus we are sampling an ensemble with a fixed average slope. 
(Our moves {\it do} access
all the microstates with a given $(\Delta_1 \h, \Delta_2 \h)$.)

We have just the same problem as if we were trying
to study the thermodynamics of an Ising model in zero field, 
but using an update which conserves spin. 
If we are in a paramagnetic state, we  know that in the
thermodynamic limit the magnetization is zero, and so we obtain
the right results if we adopt initial conditions with zero total spin;
however, we would have problems below the symmetry-breaking temperature.

Here, we believe that the thermodynamic state has zero mean height
slope (whether the heights are rough or smooth). 
Thus we require choose initial conditions with $\Delta_i \h \equiv 0$.
(This is possible when $L$ is even.)
This also means that $\z(\x)$ is single-valued and we can 
perform a Fourier transform without needing to subtract off
the average slope $\nabla\h$. 

The update rule of the simple dimer model is shown in Fig.~2(a). 
Since each color of dimer must maintain a complete covering, 
every update move of the two-color dimer models is built from
the update in Fig.~2(a): either one color is updated (as in Fig.~2(b)), 
or both colors are updated at the same time (as in Fig.~2(c) and 2(d)). 

Given a generic free energy of form (2.1), standard arguments [19] would suggest
a coarse-grained description of the Monte Carlo dynamics would be
   $$ d h_\q (t)/dt = \zeta_\q - \Gamma K |\q|^2 h_\q(t) 
   \eqno(3.1)$$
where $\Gamma$ is a kinetic coefficient and the random noise satisfies 
$\langle \zeta_\q(t) \zeta_{\q'}(t') = 2 \Gamma \delta_{\q, \q'} \delta (t-t')$.
Thus the relaxation time is expected to be wavevector-dependent as
$\tau(\q) \sim |\q|^{-2}$ [11].

\ihead {3.1.1 Dimer-loop model}

The configurations were represented in the machine as 
configurations of $\zc$. As noted above in Sec. 2.3.2,
the values of $\zsix$
may be (and were) reconstructed uniquely from those of $\zc$. 

The update moves we used correspond to 
the two rearrangements illustrated in Fig.~2(b,c) in terms of $\zc$
(or the moves related to them by symmetry). 
Both rules are necessary in order to access all the microstates. 
Notice that the move in Fig.~2(c) has no effect on the $\zsix$ coordinate.

For our simulations of the BCSOS model, the update move looks the
same as Fig.~2(c).

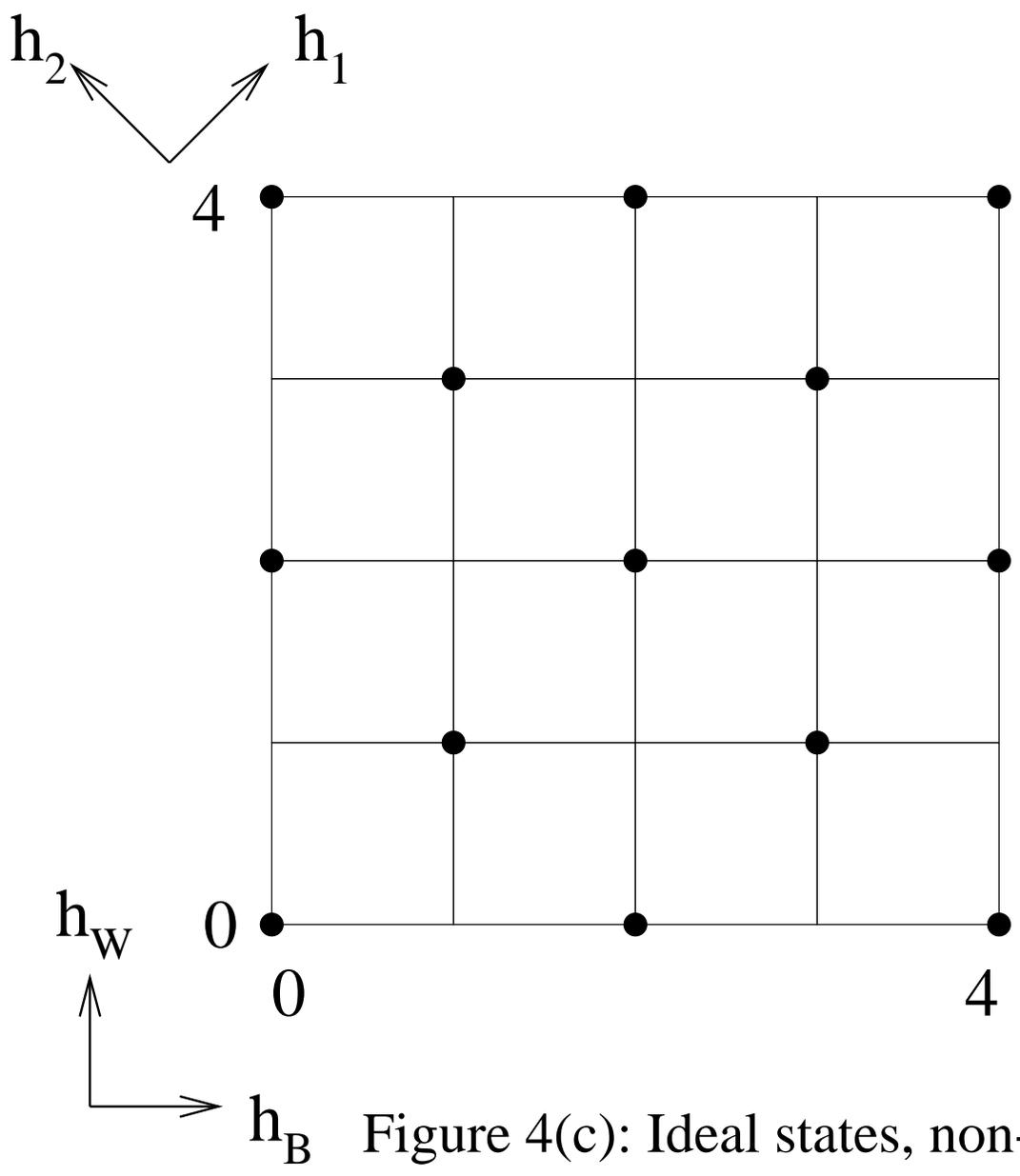
\ihead {3.1.2 Non-crossing dimer model}

The configurations were represented as patterns of $(\zB, \zW)$. 
In order to ensure detailed balance for the update of the 
noncrossing-dimer model, 
we choose at random a plaquette of the black-dimer lattice and
one of the four 
plaquettes of the dual (white-dimer) lattice that overlap 
the first plaquette;
we check whether a rearrangement involving these plaquettes
is possible, and if so we carry it out. 

Including the move shown in Fig.~2(e)
would have speeded up the simulation, but we did {\it not} implement it.
We think that the Fig.~2(d) move, by itself, in time can access 
every configuration that the Fig.~2(e) move can access.

A somewhat worrisome aspect of our update move is that it is not fully ergodic. 
Namely, if we let $M_B$ ($M_W$) be the number of black(white) dimers
oriented in the $x$ direction, then $M_B-M_W$ is {\it conserved} (by either
of the possible update moves in Fig. 2). 
However, we do not think this invalidates the results. 
The thermodynamic limit is surely dominated by $(M_B-M_W)/N$ =0, 
and our simulations have $M_B-M_W=0$ (or very nearly zero, in the 
``roof'' initial condition). Thus our ensemble, restricted by the
conservation law,  is related to the full one 
much as a microcanonical ensemble is related to a canonical one; 
such differences are usually irrelevant in the thermodynamic limit. 

A more serious criticism is that there exist certain configurations, 
(having zero mean height gradient) which contain {\it no} examples
of the updatable configuration, Fig.~2(d). A portion of a periodic
pattern of this sort is shown in Fig.~2(f). 
This means that we are in fact simulating
a modified ensemble, in which such microstates are not included. However, 
such states are a vanishing fraction of the total ensemble, and do not
matter in the thermodynamic limit. 
(If one takes the state in Fig.~2(f), 
and changes it in just one place, this introduces
an example of the Fig.~2(d) pattern; then, starting from that place, 
it is possible by iteration of the update move to 
reach the canonical flat state shown in Fig. 3(d).) 

\bhead {3.2 Simulation protocol}

One ``sweep'' consists of one random update per site. 
At sampling intervals of $n_{s}$ sweeps, we evaluate the Fourier tranform, 
and accumulate the results in sums of $|\tz(\pp)|^2$, where
    $$\tz(\pp) \equiv  {1\over{\sqrt N}} \sum _\x e^{i\pp\cdot\x} \z(\x)
   \eqno (3.2)$$
(The correlation time for the shortest wavevectors 
is long enough that sampling at shorter intervals would be
redundant.)
For both models, 
$\tz(\pp)$ was evaluated for all $\pp$ using the fast-Fourier transform (FFT).

In the case of the noncrossing-dimer model, we 
did not (and cannot!) define $\z(\x)$, as
$z_B(\x)$ and $z_W(\x)$
are defined on different sublattices of $\{ \x \}$. 
To implement to (2.19), we define
${\tilde z}_1 (\pp)$ and ${\tilde z}_2 (\pp)$ by
    $$[\pm {\tilde z}_B(\p)
   + {\tilde z}_W(\p)] /{\sqrt 2}
   \eqno (3.3)$$
Here ${\tilde z}_W(\p)]$ is actually computed by taking the FFT  of 
the function $z_W(\x+[1/2,1/2])$ and multiplying 
by the $e^{i\p\cdot(1/2,1/2)}$.
\footnote * {Notice that, by (3.3), ${\tilde z}_1(\p)$ is 
{\it not} periodic modulo the Brillouin zone, but only modulo a
doubled zone, so 
e.g., ${\tilde z}_1(\pi, p_2)\neq {\tilde z}_2(-\pi, p_2)$. The correct 
relationship is that ${\tilde z}_1(\p+(2\pi,0))= {\tilde z}_2(\p)$, 
and similarly ${\tilde z}_2(\p+(0,2\pi))= {\tilde z}_1(\p)$.}

Data was typically sampled (taking the FFT) once every 100 sweeps.
On an IBM RISC-6000/320 workstation, each dimer-loop 
run of $1.4\times 10^6$ sweeps on a $64\times 64$ 
lattice took $\sim 150$ hours, and  
each noncrossing-dimer run
of $5\times 10^5$ sweeps on a $32 \times 32$ lattice 
took $\sim 10$ hours. 
The results reported here are averaged from 
$\sim 10$ runs for each model studied. (For the test models,
simple dimers and BCSOS, much less computing effort was necessary).

To provide a check on equilibration, 
two kinds of initial condition  were used. 
\footnote \dag {The data averages reported include the entire simulation, so
the run time should be much larger than the equilibration time in order
to minimize systematic errors due to the initial configurations.}
The first is an ideal state, so that the heights are 
as flat as possible (see Fig.~3(a,b,c) again). 
All the Fourier components  $\tz(\pp)$ are initially zero, 
except at $\pp=(\pi, 0)$, $(0,\pi)$, and $(\pi, \pi)$;
typically $|\tz(\pp)|^2$ grows with time. 
The second initial condition is a ``roof'' pattern of heights, 
which is illustrated in Fig.~3(d) for the 
case of the simple-dimer model. In this case the excursion of 
$\z(\x)$ from its mean is $O(L)$, much larger than in equilibrium 
where it is $O(\sqrt{\ln L})$, and farther from equilibrium
than the ideal state. 
If the height gradient of the ``roof'' runs in the $x_1$ direction, 
then $\tz(p, 0)$ is
initially large and different in magnitude from $\tz (0, p)$.
With time, the mean-squared $\tz(p,0)$ decreases and
mean-squared $\tz(0,p)$ increases until they are equal,  
as required by symmetry. 
This was our diagnostic that equilibrium was reached in
the time allotted.
The initial difference between ``roof'' and ``flat'' initial conditions
is most extreme at the smallest wavevectors, which 
by (3.1) are the slowest relaxing ones;
thus our diagnostic provides a sensitive test 
whether our run time is adequate. 
(This is further commented in Sec. 3.3, below).

To test our method to extract elastic constants, 
we also tested {\it exactly solved} 
models with height space dimension $d\pe=1$
and with similar height mappings. 
For the dimer-loop model, the BCSOS model was our test model;
for the noncrossing-dimer model, the simple dimer model 
was the test model. 

\bhead {3.3 Results}

Our fitting procedure was to perform a linear-least-squares fit
of $\la \la |\tz_i(\pp)|^2 \ra ^{-1} $ to a polynomial in $(p_1,p_2)$ 
with all quadratic and quartic terms, using
data from a disk, roughly $|\pp| < 0.15\pi $. 
(If the quartic correction terms are omitted, the fitted 
elastic constants have systematic errors of $>1\%$.)

The data (for selected wavevectors) are shown in Figs. 5-8
for the four models (BSCOS, simple dimers, dimer-loop,  and
noncrossing-dimer).
Notice that Figs. 5-8 do not include all 
the data used for our fits: for clarity, we have shown 
only one-dimensional cuts through reciprocal space. 
However, where two values ought to be identical 
according to the symmetries of the model, we have plotted both of them. 
We premultiply by $P(\p)^2 \equiv 2(2-\cos p_x -\cos p_y)$, 
a Brillouin-zone adapted version of $|\p|^2$,  
so that the plots should asymptote to a constant as $\p \to 0$, 
if they follow
the expected behavior for a rough interface.\footnote * {In the case of
  the simple dimer model, replacing $|\p|^2$ by $P(\p)^2$ made
  a visible difference: most of the variation of $|\p|^2 \la |\tz(\p)|^2\ra$ 
  with $\p$ was cancelled.}

In the non-crossing dimer plot, one of the 
data points for the shortest wavevector ($|\p|=2\pi/32$) 
is too low (compared to the line through the other data points),
while the data points for the other symmetry-related wavevectors
of the same length are about right.
This is due solely to 
the Fourier component ${\tilde z}_1(\pi/16,0)$ being spuriously large
(by a factor of 1.5) in the data sets using the ``roof'' initial condition.
That shows that the 
equilibration time we allowed was, in fact, {\it not}
longer than the longest relaxation time
(however, all other wavevectors appear to be equilibrated). 
Presumably the same explanation applies to the similar (but smaller) 
discrepancies visible in some of the other plots.


The elastic constants extracted from the best fits are shown 
in Table~I. The error which we choose to quote is  taken
as twice the statistical error, to allow for systematic errors. 
(The statistical errors
were calculated from the variance of the fitted elastic constants 
from at least four independent runs of the same model.)
For the new models, we indicate conjectures (marked by ``?'' in
the Tables), assuming that
the stiffnesses are simple rational multiples of $\pi$
(which would be the condition for having rational critical exponents). 

Table~II shows the predicted correlation exponents
when the (known or conjectured) exact stiffnesses in Table~I are
inserted into the theoretical formulas (2.8) and (2.9) 
(or their generalizations (2.24), (2.25), (2.31), and (2.32) in Sec. 2.5). 
Recall that $\eta(\G\pe)$ are exponents for
local operators, e.g. the dimer-dimer correlation, while 
$\eta(\b)$ are for defect operators. 

\bbhead {4. DISCUSSION OF RESULTS}

In this section, we review our expectations 
for a general height model, in the light of Sec.~2.1, and
then try to make sense of the prominent features of the results.

\bhead {4.1 General expectations for behavior of $\la|\tz(\p)|^2\ra$}

The prime expectation is a consequence of coarse-graining
(Sec. 2.1). 
Near $\pp=0$, i.e. averaged over long wavelengths, $\z(\pp) \approx \h(\pp)$. 
Thus if the height field is in a ``rough'' phase , eq. (2.3) implies 
  $$\la |{\tilde z}(\pp)|^2 \ra ^{-1}  \sim K |\pp|^2
  \eqno (4.1)$$
in a one-dimensional height model; for $d\pe=2$, (4.1) has its 
obvious generalizations in terms of the elastic theory of Sec.~2.4.

On the other hand, our usual picture of a ``smooth'' phase
is that the system is in an ideal state on the majority of the sites. 
The fluctuations in equilibrium consist of individual sites, 
or very small domains, on which the heights deviate from the ideal state. 
Since these fluctuations are local and do not overlap, 
they ought to be independent. 
Hence their Fourier transform should have a white-noise spectrum, 
   $$\la |{\tilde z}(\pp)|^2 \ra \sim  const 
   \eqno (4.2)$$
as $\pp \to \O$.

It should be noted that the general definition of
a ``smooth'' phase in an SOS model is that
the net height variance 
   $$W^2 \equiv \la \z(\x)^2 \ra - \la \z(\x) \ra ^2 
   \eqno (4.3)$$
is finite in the thermodynamic limit. 
Thus, smoothness means that the system undergoes a symmetry-breaking
in which a particular mean height is picked out with long-ranged order. 
Writing the height variance as
     $$ W^2 = \sum _{\pp\neq 0} \la |{\tilde z}(\pp)|^2 \ra. 
    \eqno (4.4)$$
it can be seen that on insertion into (4.4), the smooth behavior
(4.2) indeed gives a convergent sum 
while the rough behavior (4.1) gives the well-known logarithmic divergence, 
   $$ W^2= {\rm const} + {1\over {2\pi K}}\ln L
    \eqno (4.5)$$ 
since the sum (4.4) is cut off at the smallest $\pp$ which is of order $1/L$. 

A generic feature in height model data is the presence 
of zone-boundary singularities of $\la |\z(\p)|^2 \ra$ for
\def \edge {^{(1)}}
\def \cor  {^{(2)}}
$\p$ near $\Q\edge=(\pi,0)$  and $\Q\cor=(\pi,\pi)$.
This is a consequence of the fact that $\z$ is not uniform
even in an ideal state; rather $\z-\hbarr$ is modulated at wavevectors 
$\Q^{(i)}$. The amplitude of modulation itself is a periodic function
of the heights, thus it is an operator of form 
${\tilde O}(\h) \sim \exp (i\G^{(i)} \cdot \h) $,  
implying power-law correlations as explained in Sec. 1.
Following this through (Ref. [10] gives more details for the case
of the 4-coloring model), 
we predict 
    $$ \langle \tz(\p) |^2 \rangle \sim |\p-\Q^{(i)}|^{-(2-\eta^{(i)})}
    \eqno (4.6)$$
where $\eta^{(i)} = \eta(\G^{(i)})$. 

In the case of the simple dimer model, 
examination of the modulations gives $G^{(1)}= 2\pi/4$ and
$G^{(2)}=2\pi/2$, implying (see Table II) that 
$\eta^{(1)}=1/2$ and $\eta^{(2)}=2$. 
Examination of our data from this model shows that 
the fluctuations indeed have a small peak at $(\pi,0)$ and 
are constant near $(\pi,\pi)$. 
Such zone-boundary peaks have also been seen in simulations 
of other height models [10,11].

In the discussion of each model, 
we will check the plausibility of the behavior and numerical
value of the stiffness constant  by considering the models
as modifications of exactly solved models. 
In particular, we may define generalized
``ghost'' versions of the two-color dimer models 
by discarding the hard-core exclusion between black and white dimers.
Instead we might include a weight factor $e^{-u}$ in the partition function
for each such violation. 
Thus the $u=\infty$ case is the models we simulated, 
and the $u=0$ case is the noninteracting limit
consisting of two copies of the simple dimer model.

Presumably the stiffness constants are monotonic functions of $u$. 
Since the exclusion rules still permit flat ``ideal'' states, but 
permit fewer ways of fluctuating away from them, we anticipate
that the stiffness constants increase as we turn on $u$.
If so, then the elastic constants for the noninteracting cases,
which may be found trivially by applying the changes of variables
(2.17) and (2.18), supply a lower bound on the expected results. 

\bhead {4.2 Dimer-loop model}

The dimer-loop model data 
could be fit to $\la |{\tilde \zsix}(\pp)|^2 \ra \sim 1/(\Ksix |\p|^2)$ 
as expected for the ``rough'' behavior (4.1). 
However, it is evident from Fig. 7(a)
that (after excluding the smallest $\p$ value, see Sec. 3.3), 
these data deviate more strongly at small $\p$ in the other models.
They might be fit to a different exponent, roughly
$\langle |\z_1(\p)|^2 \rangle \sim |\p|^{-1.9}$, but this is so close to 
$|\p|^{-2}$ that the latter seems more plausible.
Perhaps the deviation can be blamed on 
coupling to the anomalous smooth component $z_2$ (see below). 
As expected, $\la |{\tilde \zsix}(\pp)|^2 \ra$ also shows slight maxima 
at the zone-boundary points $\Q^{(1)}$ and $\Q^{(2)}$.) 

However, as is obvious from Fig.~7(b),
the second height component  has a different behavior:
   $$\la |{\tilde \zc}(\pp)|^2 \ra \sim |\pp|^{-(2-\eta^{(0)})}
    \eqno (4.7)$$ 
with $2-\eta^{(0)} \approx 1$,  very roughly.
(The apparent exponent varies from $\sim 0.6$ at the smallest $|\p|$
to $\sim 1.2$ at larger $|\p|$.)
The fact that $\eta^{(0)}>0$ implies the $\zc$ component is ``smooth'' in
the sense that the its variance, $W_2^2$, is finite.

Thus we are led to an approximate picture in which $\zc(\x)$ is 
an ideal state configuration,  on which bounded fluctuations are
superposed.
(It is quite possible to have anomalous power-law correlations,
yet have strictly bounded height fluctuations.)
Indeed, examination of Fig. 1 (b) shows that
$\zc$ deviates at most $\pm 2$ from the ideal state; furthermore
the mean $\zc$ is 7.51, compared with 7.5 in the ideal state.
The same picture predicts that 
$\la |{\tilde \zc}|^2\ra$ has a zone-boundary peak at 
$\Q_1=(\pi,0)$ approximating a $\delta$-function, which is 
indeed seen in the data. 

\ihead { 4.1.1 The anomalous smoothness of $\zc$}

The smoothness of $\zc$ is anomalous in that (4.7) is quite 
different from (4.2), the behavior for a generic ``smooth'' phase.
The probable explanation is that $\zc$ deviations 
experience a power-law effective coupling, mediated by  
the critical $\zsix$ fluctuations.

As an approximation, let us take an ansatz that $\zc$ is constrained to take on
only two values, say 0 and 1.
In the partially coarse-grained picture (see Sec. 2.2), 
the system consists entirely of domains with $\hc=1/2$, 
while $\hsix$ can still take any value $m+1/2$. 
However, it can be checked that each domain wall has a net excess
$\delta \zc = \pm 1/4$  per lattice constant in the $x_1$ or $x_2$ direction.
Let us say $\hsix=m$ along the domain wall separating states with 
$\hsix=m-1/2$ and $\hsix=m+1/2$. It turns out that the sign of the excess is 
independent e.g. of the wall orientation; it is a function 
only of m (or $\hc$) with period 4,  taking values in 
the repeating sequence $1/4, 1/4, -1/4, -1/4$.
This can be written  
   $$\delta \zc \simeq {1\over {2\sqrt 2}} \cos {\pi\over 2}(\hc-{1\over2}).
   \eqno(4.8)$$
Thus we have put the operator $\delta \zc$  into the form (2.6)
with $\G\pe=(\pi/2,0)$. Consequently, using  (2.8) and the
conjectured exact value of $\Ksix$ from Table I, we obtain
$\eta^{(0)} = \eta(\G\pe) \approx 1/2$. 
(The data (4.7), however, definitely point to a smaller value of $\eta^{(0)}$.)

\ihead {4.1.2 Comparison to exactly solved models}

First let us forget the $\zc$ component (since it is smooth anyhow)
and consider the uncolored FPL model introduced in Sec. 2.2, 
with stiffness $K_{FPL}(n)$.
The $n=1$ case of the FPL is the equal-weighted
a six-vertex or ``ice'' model, which in the height representation
is equivalent to the (exactly solved) BCSOS model.
Now, increasing the loop fugacity $n$ from 1 favors configurations which have
many small loops, and are thus flatter in the height represenation. 
(The state with the maximum number of loops is the
one in Fig.~3(c), which also has maximum flatness.) 
Thus $K_{FPL}(n)$ should increase with $n$, 
and indeed Table I shows 
that $K_1\equiv K_{FPL}(2)$ 
is larger than the stiffness 
of the (equal-weighted) BCSOS model 
$K=K_{FPL}(1)=\pi/6$.

For an independent comparison to an exactly solved model, consider
the ``ghost'' dimer-loop model, in its noninteracting limit. 
This has elastic constants $\Ksix^{(0)}=2K_0= \pi/8 \approx 0.3927$ and 
$\Kc^{(0)}=K_0= \pi/2$. 
It can be checked, using Fig.~4(b), that the reciprocal lattice
vector for locking in the $\zc$ direction is
$\G_{lock}^{(2)}=(0,2\pi)$ and hence $\eta(\G_{lock}^{(2)})=4$:
thus the ``ghost'' model is exactly marginal.
Increasing $\Kc$, 
by turning on the dimer exclusion $u$ infinitesimally, 
would reduce $\eta(\G_{lock}^{(2)})$, and
consequently by (2.10) $\hc$ should immediately lock (unroughen). 

On the other hand, returning to the FPL representation, it has been observed
recently that in loop models, $\eta(\G_{lock})$ is typically marginal, 
for the height component whose contours form the loops [2].
(Such models include the honeycomb lattice FPL model 
[7,8], and the 4-coloring model on the square lattice [10].)
This criterion would predict a finite value $\Kc=\pi/2$  
for the dimer-loop model
(presumably $\Kc$ would stay constant for {\it all} $u$ 
in the ``ghost'' model; only $\Ksix$ would increase as the dimer exclusion is
gradually turned on.)
Our data (Fig.~7(b)) clearly exclude this possibility;
clearly $z_1$ is not marginal either,
since $\eta(\G_{lock}) = \eta((2\pi,0))\approx 8$.

\bhead {4.3 Noncrossing-dimer model}

After converting (2.21) and (2.22) from $\qq$ coordinates back to 
$\pp$ coordinates, we obtain 
    $$ \la |z_1(\pp)|^2 \ra^{-1}  = 
\half (\Kp+\Km) |\pp|^2  + \half (\Kp-\Km)  2p_1p_2 
    \eqno (4.9)$$
near $\pp=\O$, 
and the same for $\la |z_2(\pp)|^2\ra$ except for
a change of sign in the second term. 
The lack of isotropy seen in Fig.~8 is consistent with (4.9).

The difference in Table~I between the measured $\Kp$ 
and our conjectured exact value $\pi/12$ is significant.
We have included this conjecture, nevertheless, for two reasons.
Firstly, fits using a somewhat larger portion of $\p$ space 
yielded systematically higher values of the elastic constants, 
suggesting that our best fit value is still a bit too high;
secondly, this conjecture implies a rational value of $(\Kp/\Km)^{1/2}$, so that
the geometric mean $\Kbar$ (see Sec. 2.5) can also be a 
rational multiple of $\pi$,
which in turn (by (2.31) and (2.32)) implies {\it rational} critical exponents. 
Our conjectured value is 
    $$\Kbar= \pi/9 
    \eqno (4.10) $$

The data show zone-boundary peaks of $\la|\tz(\p)|^2\ra$ 
near $\Q^{(1)}=(\pi,0)$;
the existence of peaks suggests that $\eta^{(1)}<2$, 
however the data  are too crude to estimate the exponent quantitatively. 
On the other hand, 
the theory of eq. (4.6) would predict $\eta^{(1)}=\eta(\pi/2, \pi/2)= 9/4$. 

Near the zone corner $\Q^{(2)}=(\pi,\pi)$, 
more complicated behavior is observed.
Approaching along the (1,1) direction, 
the $\la|{\tilde z}_1(\p)|^2\ra$ component  {\it vanishes}
as $\p \to \Q^{(2)}$,  roughly as $|\p-\Q_2|^2$; in accord with the
symmetry here, $\la|{\tilde z}_2(\p)|^2\ra$ behaves the same along the
(1,-1) directions. 
(Approaching $\Q^{(2)}$ along the orthogonal directions, 
a constant limit is observed, but this is really an indpendent 
wavevector; see the footnote in Sec. 3.2).

To understand what is going on at $\Q^{(2)}$, 
note that ${\tilde z}_1(\Q^{(2)}) = 2(M_B-M_W)$, the excess of 
dimers of one color in a given orientation.
Evidently, the long-wavelength fluctuations of this quantity 
are divergent. Possibly this behavior is an artifact of the
dynamics we used, which (as noted earlier) conserve $M_B-M_W$. 
But if that is the case, it cannot simply be a memory effect 
of the initial configuration
(since that has $M_B-M_W \approx 0$). 
It is equally hard to blame the divergence on random fluctuations 
which develop during the runs 
(and then persist because the conservation slows down the dynamics at those 
wavevectors): the divergent part does not differ from run to
run, as would be expected from this explanation. 
An alternative hypothesis is that the excess $M_B-M_W$ (or, more precisely, the
corresponding local density of the excess) 
functions like a third height variable
in the free energy (thus explaining the inverse-square power of the divergence).
However, it cannot be a height variable in the usual sense,
since the system is not invariant under shifts of the excess.
Thus, at present we have no convincing explanation of the divergence
at $\Q^{(2)}$. 

\ihead {Comparison to exactly solved models}

The noninteracting ``ghost'' dimer model has 
$\Kp(0) = \Km(0) = K_0  \approx 0.3927$. 
The value for $\Kp$ in Table II  is, as expected, 
greater than $\Kp(0)$, however $\Km$ is {\it smaller} than $\Km(0)$.

\bbhead {5. CONCLUSIONS }

We have defined two new dimer models which have two-dimensional
height representations as worked out in Sec. 2.2 and  2.3. 
(A third new height model presented here is the 
Ising model mentioned in Sec.~4.2, which was not simulated). 
By accumulating the mean-squared height fluctuations of each Fourier
mode as analyzed from a Monte Carlo simulation, we have 
extracted estimates of the stiffness constants which are
accurate to $\sim 1\%$, using modest system sizes (Sec. 4 and
Table~I). 
Via the Coulomb-gas-like notions in Sec. 2.1 and 2.5, 
the stiffnesses imply critical exponents (see Table~II). 
The latter show that our models belong to new universality classes, 
and (for the dimer-loop model) the exponents are 
suggestively close to being simple rational numbers.
Each of our models has an anomalous feature: 
in the dimer-loop model the $\zc$ height component is smooth yet has
critical correlations (Sec.~4.2),  
and in the noncrossing-dimer model correlation functions
are anisotropic even in the critical limit (Sec. 2.5). 

In this work, we have tested a new and better Monte Carlo technique for
evaluating  critical exponents. 
The results of these simulations (and those of other height models 
[10,11]) clearly show
that the most efficient way to extract  $K$ is 
to measure the Gaussian height fluctuations 
and use (2.3) with (2.8) and (2.9) (or their generalizations)
to infer the exponents, rather than 
to measure correlation functions directly.
Unfortunately, this approach is applicable only to height models 
(and even spin models which have height
representations at $T=0$ generally do not at $T>0$.) 

We can see this comparison a bit more sharply, because 
(as discussed in Sec.~4.1) the measured, Fourier-transformed
height fluctuations $\la |\tz(\p)|^2\ra$
in fact contain not only the peak around $\p=\O$
from the long wavelength modes, 
but also peaks around special points on the zone boundary,
proportional to the structure factor of a local operator.
In other words, as a side-effect we have directly measured 
(the Fourier transform of) the correlation function of that local operator;
thus the singular powers $2-\eta^{(1)}$ and $2-\eta^{(2)}$
found around the zone boundary peaks 
are the best estimates of the exponents that could be obtained by the
standard approach.
But in fact those peaks were not strong enough to extract 
meaningful values of the exponents; a larger system size would be necessary. 
Since the standard analysis using the operator
correlation functions uses the {\it same} data that enter the 
``height-fluctuation'' analysis, 
this gives a fair measure of the superiority of the latter approach.

It is interesting to compare our methods to another 
older literature, that of SOS models simulations. 
These simulations [20] have frequently determined $K$ by 
fitting the system size dependence of the net height variance $W^2$ 
to eq. (4.5).
Such a method throws away much of the information in the smallest 
$\pp$ wavevectors by integrating over them.
In other cases [21] authors have evaluated the correlation function
$C(\x)$ fitting to eq. (2.5).
This correlation function, in principle, has the same information
as $\la |{\tilde z(\pp)}|^2\ra$, but
it does not exclude the large-$\pp$ modes (with
their systematic errors) quite as cleanly.
In either of the two older methods, 
$K$ would be  extracted from a linear fit to the logarithmic dependence. 
Either of those methods would work poorly for our data, 
due to the limited range of $\ln r$ offered with system size $L=32$. 
With the Fourier method, we can already get the correct stiffness
to $\sim 10^{-3}$ (in the simple dimer case, which seems
especially favorable).

The dimer-loop model  is a good candidate
for a Bethe-ansatz exact solution, since
it maps to the square-lattice fully-packed loop (FPL) model
with fugacity $n=2$, while the honeycomb FPL model
is soluble [22]. 

It is also interesting to note that the dimer-loop model
is a limiting case of the family of ``loop-gas'' models, which
also arise from the superposition of two dimer coverings.
Namely, we this loop-gas has a
fugacity for ordinary loops $y=2$, and a fugacity 
for loops of length 2 (corresponding to superposed black and white
dimers) is $x=0$. [23]
(The original case of the loop gas had $y=4$ and $x=2$, 
and described the correlation functions of 
the nearest-neighbor resonating-valence bond variational wavefunction for 
the spin-$1/2$ antiferromagnet on the square lattice.)

Another promising direction would be to study the ``ghost'' versions
of the two-color dimer models 
by a Kosterlitz-Thouless type of renormalization group, 
e.g. to determine how the coupling between colors $u$ 
renormalizes when it is small.

\bhead {Acknowledgments}
We would like to thank J. K. Burton, J. Kondev, and C. Zeng
for discussions. C.L.H. was supported by NSF Grant DMR-9214943.
and R.R. by  the David and Lucile Packard Foundation.

\vfill\eject
\bhead{REFERENCES}

[1] R. Liebmann, 1986, 
{\it Statistical Physics of Periodically Frustrated Ising Systems}
(Springer Lecture Notes on Physics No. 251).

[2] 
J. Kondev and C. L. Henley, 
Nuc. Phys. B 464, 540 (1996)
[cond-mat/9511102];
see also J. Kondev and
C. L. Henley, Phys. Rev. Lett. 74, 4580 (1995), and
J. Kondev, in {\it Proc. Symposium 
on Exactly Soluble Models in Statistical Mechanics}, 
(Northeastern Univ., March 1996), ed. C.~King and F.~Y.~Wu. 

[3]
C. L. Henley, unpublished.

[4] See e.g. D. R. Nelson in 
{\it Phase Transitions and Critical Phenomena}.
ed. C. Domb and J. L. Lebowitz (?)
(Academic Press, London, 1983), vol.~7.

[5] H. W. J. Bl\"ote and H. J. Hilhorst,
J. Phys. A 15, L631 (1982);
B. Nienhuis, H. J. Hilhorst, and H. W. J. Bl\"ote,
J. Phys. A 17, 3559 (1984).

[6] D. A. Huse and A. D. Rutenberg, 
Phys. Rev. B45, 7536 (1992).

[7] 
H. W. J. Bl\"ote and B. Nienhuis, 
Phys. Rev. Lett. 72, 1372 (1994).

[9] N. Read 1992,  reported in Kagom\'e workshop (Jan. 1992), 
and unpublished.

[10] J. Kondev and C. L. Henley, 
Phys. Rev. B 52, 6628 (1995).

[11] C. Zeng and C. L. Henley, unpublished. 

[12] J. K. Burton and C. L. Henley,  preprint, 
to be submitted to J. Phys. A (1996). 

[13] J. Kondev and C. L. Henley, 
Phys. Rev. Lett. 74, 4580 (1995)

[14] B. Nienhuis in 
{\it Phase Transitions and Critical Phenomena}, 
ed. C. Domb and J. L. Lebowitz
(Academic Press, London, 1987), vol. 11.

[15]
H. N. V. Temperley and M. E. Fisher,
Phil. Mag. 6, 1061 (1961);
P. W. Kasteleyn, Physica 27, 1209 (1961);
M. E. Fisher and J. Stephenson, 
Phys. Rev. 132, 1411 (1963);
J. Stephenson, J. Math. Phys. 5, 1009 (1964).

[16] See E. H. Lieb and F. Y. Wu 1972,
in {\it Phase Transitions and Critical Phenomena},
ed. C. Domb and M. S. Green, 
(Academic Press, 1972), vol. 1.

[17] W. Zheng and S. Sachdev, Phys. Rev. B 40, 2704 (1989);
L. S. Levitov,  Phys. Rev. Lett. 64, 92 (1990);
L. B. Ioffe and A. I. Larkin, 
Phys. Rev. B 40, 6941 (1989).

[18] H. van Beijeren, Phys. Rev. Lett. 38, 993 (1977).

[19] C. L. Henley, submitted to J. Stat. Phys.

[20] 
H. G. Evertz, M. Hasenbusch, M. Marcu, K. Pinn, and S. Solomon, 
J. Phys. (France) I 1, 1669 (1991).

[21] W. J. Shugard, J. D. Weeks, and G. H. Gilmer, 
Phys. Rev. Lett. 41, 1399 (1978);

[22] M. T. Batchelor, J. Suzuki, and C. M. Yung,
Phys. Rev. Lett. 73, 2646 (1994).

[23] B. Sutherland, Phys. Rev. B 38, 6855 (1988), and unpublished.

\vfill
\eject

\bbhead {FIGURE CAPTIONS}

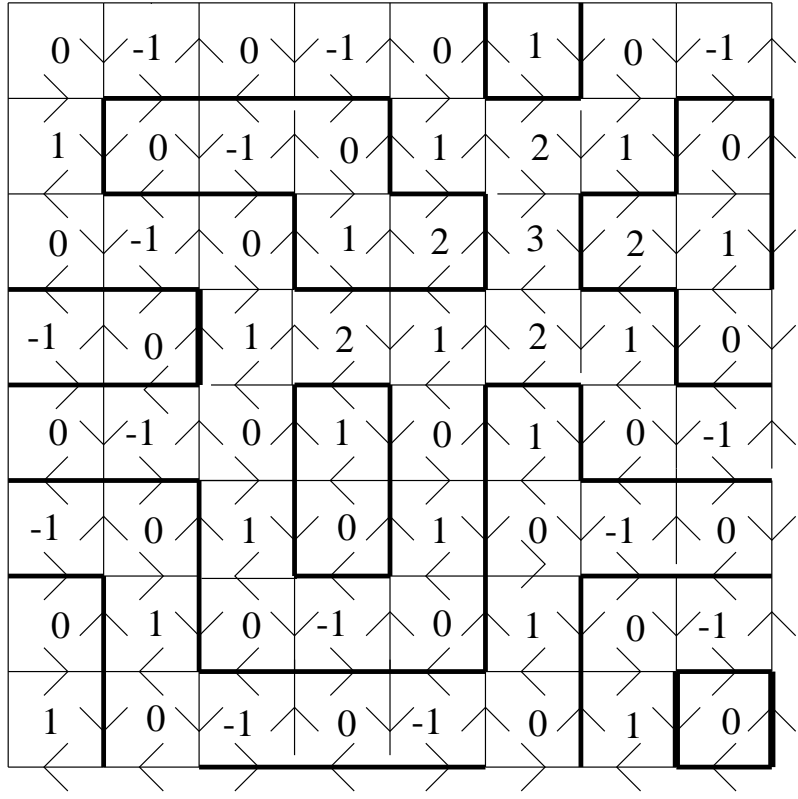
Fig. 1. Sample configurations of the models 
from simulations  of $8\times 8$ lattice:
(a). Simple dimer model, heights $z(\x)$ marked;
(b). dimer-loop model, with $\zc(\x)$ marked;
(c). the same loops (dimer colors not distinguished) 
with arrows showing the mapping to the 6-vertex model, and
height component $\zsix$ marked;
(d). noncrossing-dimer model.

\medskip

Fig. 2.  Rearrangements 
between two microstates; the first four form the basis of Monte Carlo
updates (see Sec.~3).  
(a). Simple dimer model, showing the heights $z(\x)$. (b,c) 
Two-color dimer-loop covering, showing heights $\zc(\x)$. 
(d,e). Two-color noncrossing-dimers covering. 
(f). Fragment of a state in the noncrossing-dimer model
which cannot be updated. 
Rearrangements (a,b,d) produce the smallest possible difference
between ground states. 

\medskip

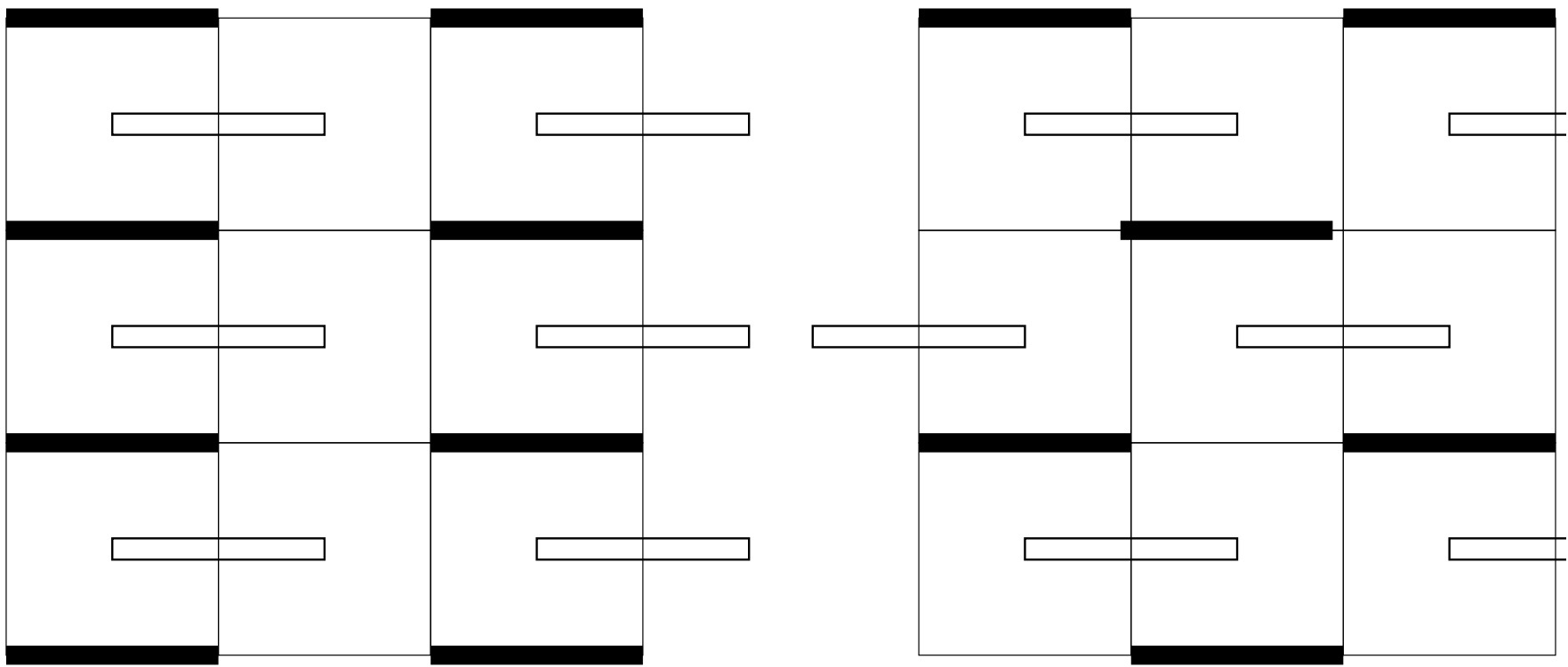
Fig. 3. Ideal states
(a). for simple dimer model,  (b) for dimer-loop model, 
(c). also for dimer-loop model (alternative candidate 
ideal  state)
and (c). for noncrossing-dimer model.
In (d), part of a ``roof'' configuration is shown for the
noncrossing-dimer model;
this is the opposite of an ideal state, since it has maximum
height gradients.

dimer-loop... when I eyeball the curves, I get 0.76+-.02, and it is
easier to believe it deviates up rather than down.
So I'll bet for 0.785.
\medskip

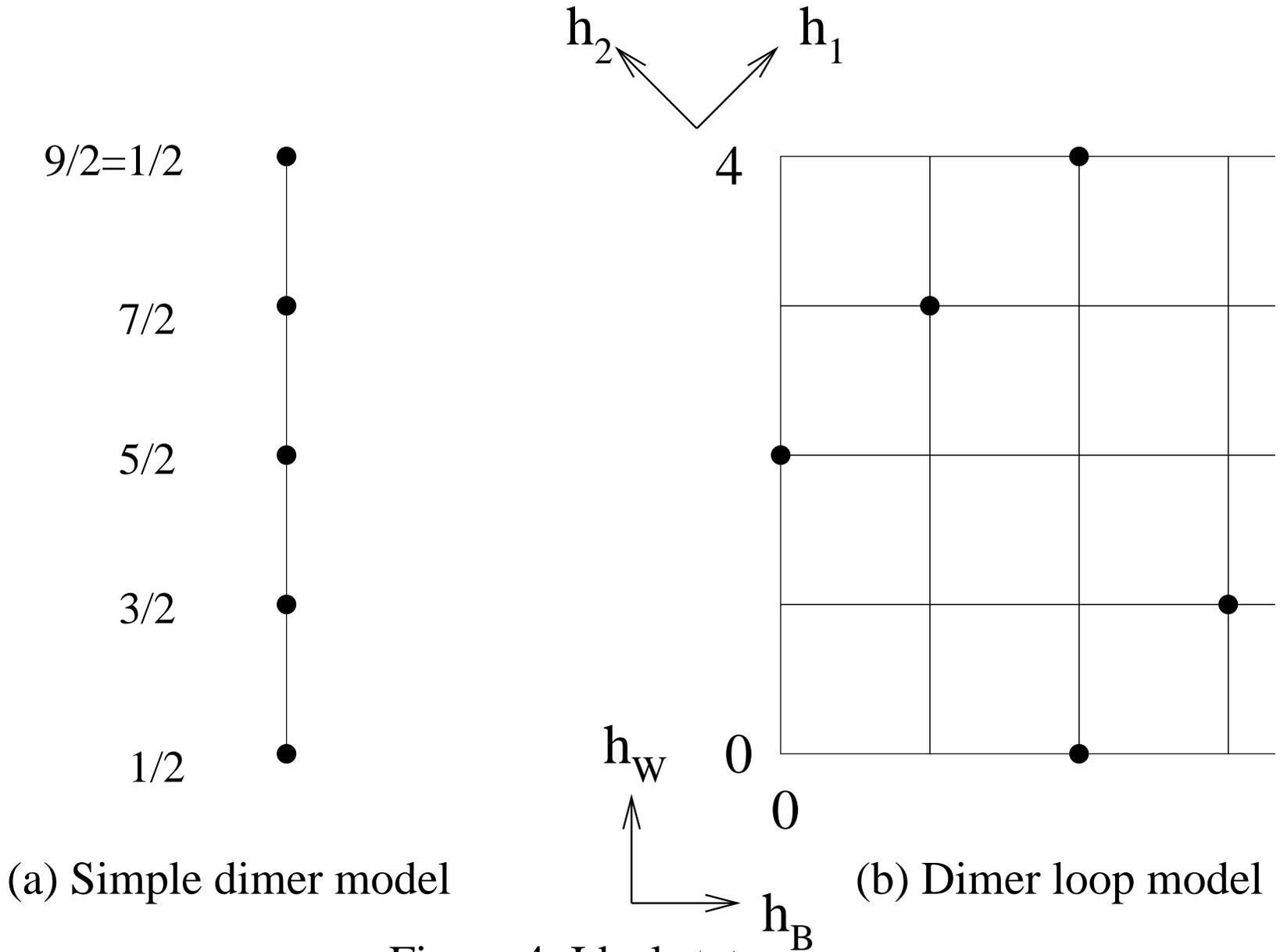
Fig. 4. Graph of ideal states in height space (heavy dots).
The dashed grid marks the ideal states of the ghost (noninteracting)
two-color dimer models.
Rotated axes are indicated  corresponding to the even and
odd combinations of $(\hB, \hW)$ which appear in the elastic
free energy (see Sec. 2.4). 
(a). Simple dimer model
(b). Dimer-loop model
(c). Noncrossing-dimer model. 

\medskip

Fig. 5.  Simple dimer model results.
$[P(\pp)^2 \la |{\tilde z}(\pp)|^2 \ra] ^{-1}$
is plotted against the squared wavevector $|\pp|^2$;
where ${\tilde z}(\pp)$ is the Fourier transform of the 
configuration of heights.  
(Here $P(\pp)^2 \approx |\pp|^2$ near $|\pp|=0$, see text). 
In light of eq. (2.3), the graph should approach
a constant near $|\pp|=0$.
In order to test the isotropy in real space, 
we show data for wavevectors $\pp$ along 
$(10)$ and $(11)$  directions by 
filled and open circles, respectively. 

\medskip

Fig. 6. BCSOS model results, using the same conventions as in Fig. 5.

\medskip

Fig. 7. Dimer-loop results
(a). Rough component $\zsix$, using the conventions of Fig.~5.
(b). Anomalously smooth component $\zc$, using log-log plot to  
check power law behavior.

\medskip

Fig. 8. Noncrossing-dimer results
Here data for wavevectors along the  $(1,-1)$ axis
must be distinguished from those along $(1,1)$ axis.
The limit at $\pp\to 0$ is expected to be a function of direction $\hat \pp$ 
in this model.

\end